\documentclass[11pt]{article}
\usepackage[left=20mm, right=20mm, top=15mm, bottom=15mm, includehead=true, includefoot=true]{geometry}

\usepackage[colorlinks=true,allcolors=black,citecolor=blue, linkcolor=blue]{hyperref}
\usepackage{listings}
\usepackage{color}
\usepackage{graphicx}
\usepackage{float}
\usepackage{graphicx}
\usepackage{natbib} 
\usepackage{url}
\usepackage{authblk} 
\usepackage{todonotes}
\usepackage{float}
\usepackage{listings}
\usepackage{rotating}
\usepackage{multirow}
\usepackage{multicol}
\usepackage{booktabs}
\usepackage{enumitem}
\usepackage{longtable}
\usepackage[parfill]{parskip} 
\usepackage{array,etoolbox}
\usepackage{setspace}
\usepackage{lineno}

\usepackage{tikz}
\usepackage[tikz]{mdframed}

\definecolor{light-gray}{gray}{0.95}
\newmdenv[%
outerlinewidth=2,%
roundcorner=10pt,%
linecolor=gray,%
backgroundcolor=light-gray,%
]{myframe}

\usepackage{helvet}

\preto\tabular{\setcounter{magicrownumbers}{0}}
\newcounter{magicrownumbers}

\usepackage{sectsty}
\sectionfont{\normalsize}
\subsectionfont{\normalsize}
\subsubsectionfont{\normalsize}
\paragraphfont{\normalsize}

\newcommand{\ie}{i.e.}
\newcommand{\eg}{e.g.}

\newcommand{\Reffig}[1]{Figure~\ref{#1}}

\newcommand{\Reftab}[1]{Table~\ref{#1}}

\definecolor{gray}{rgb}{0.4,0.4,0.4}
\definecolor{darkblue}{rgb}{0.0,0.0,0.6}
\definecolor{cyan}{rgb}{0.0,0.6,0.6}
\definecolor{mauve}{rgb}{0.58,0,0.82}

\lstdefinelanguage{XML}
{
  frame=tb,
  aboveskip=3mm,
  belowskip=3mm,
  breaklines=true,
  breakatwhitespace=true,
  morestring=[b]",
  morestring=[s]{>}{<},
  morecomment=[s]{<?}{?>},
  stringstyle=\color{darkblue},
  identifierstyle=\color{darkblue},
  keywordstyle=\color{mauve},
  morekeywords={xmlns,version,type,name,codeSpace}
}

\title{Reference study of CityGML software support: the GeoBIM benchmark 2019 --- Part II}

\author[1]{Francesca Noardo}
\author[1]{Ken Arroyo Ohori}
\author[2,3]{Filip Biljecki}
\author[4]{Claire Ellul}
\author[5]{Lars Harrie}
\author[1]{Thomas Krijnen}

\author[5]{Helen Eriksson}
\author[1]{Jordi van Liempt}
\author[6]{Maria Pla}
\author[6]{Antonio Ruiz}

\author[7]{Dean Hintz}
\author[8]{Nina Krueger}
\author[9]{Cristina Leoni}
\author[10]{Leire Leoz}
\author[11]{Diana Moraru}
\author[1]{Stelios Vitalis}
\author[8]{Philipp Willkomm}

\author[1]{Jantien Stoter}

\affil[1]{3D Geoinformation, Delft University of Technology, Delft, The Netherlands --- f.noardo@tudelft.nl, k.ohori@tudelft.nl, t.f.krijnen@tudelft.nl, j.n.h.vanliempt@student.tudelft.nl, s.vitalis@tudelft.nl, j.e.stoter@tudelft.nl}

\affil[2]{Department of Architecture, National University of Singapore, Singapore --- filip@nus.edu.sg}
\affil[3]{Department of Real Estate, National University of Singapore, Singapore}

\affil[4]{Department of Civil, Environmental and Geomatic Engineering, University College London, London, UK --- c.ellul@ucl.ac.uk}

\affil[5]{Department of Physical Geography and Ecosystem Science, Lund University, Sweden --- (lars.harrie, helen.eriksson)@nateko.lu.se}

\affil[6]{Institut Cartogr{\`a}fic i Geol{\`o}gic de Catalunya --- (maria.pla, antonio.ruiz)@icgc.cat}

\affil[7]{Safe Software, Surrey, Canada --- dean.hintz@safe.com}

\affil[8]{M.O.S.S.
Computer Grafik Systeme GmbH, Taufkirchen, Germany --- (nkrueger, pwillkomm)@moss.de}

\affil[9]{Department of Civil, Constructional and Environmental Engineering, Sapienza University of Rome, Rome --- cristina.leoni@uniroma1.it}

\affil[10]{Tracasa, Pamplona, Spain --- lleoz@tracasa.es}

\affil[11]{Ordnance Survey, United Kingdom --- diana.moraru@os.uk}

\begin{document}

\maketitle

\begin{myframe}
	
	This is the author's version of the work. 
	
	It is posted here only for personal use, not for redistribution and not for commercial use.
	
	The definitive version is published in the journal \emph{Transactions in GIS}.
	\\	
	\\
	Noardo, F., Arroyo Ohori, K., Biljecki, F., Ellul, C., Harrie, L., Krijnen, T., Eriksson, H., van Liempt, J., Pla, M., Ruiz, A., Hintz, D., Krueger, N., Leoni, C., Leoz, L., Moraru, D., Vitalis, S., Willkomm, P., Stoter, J. (2020). Reference study of CityGML software support: the GeoBIM benchmark 2019 – Part II. \emph{Transactions in GIS}.
	\\\textsc{doi}: \url{https://doi.org/10.1111/tgis.12710}
	\\
	\\
	Full details of the project: \url{https://3d.bk.tudelft.nl/projects/geobim-benchmark/}
	
\end{myframe}



 \begin{abstract}
OGC CityGML is an open standard for 3D city models intended to foster interoperability and support various applications.
However, through our practical experience and discussions with practitioners, we have noticed several problems related to the implementation of the standard and the use of standardized data.
Nevertheless, a systematic investigation of these issues has never been performed, and there is thus insufficient evidence that can be used for tackling the problems.
The GeoBIM benchmark project is aimed at finding such evidence by involving external volunteers, reporting on tools behaviour about relevant aspects (geometry, semantics, georeferencing, functionalities), analysed and described in this paper.
This study explicitly pointed out the critical points embedded in the format as an evidence base for future development.
This paper is in tandem with Part I, describing the results of the benchmark related to IFC, counterpart of CityGML within building information modelling\@.

 $ $ 

 \end{abstract}


\section{Introduction}\label{sec:intro}

Interoperability through open standards is critical for the effective re-use and exchange of data and it is essential for reciprocal integration of data having different nature.
The integration of 3D city models with Building information models (BIMs) has become a widely discussed topic in recent research.
Two open standard data models considered for accomplishing such an integration are the Open Geospatial Consortium CityGML\footnote{\url{http://www.citygmlwiki.org}} for 3D city models, and buildingSMART Industry Foundation Classes (IFC)\footnote{\url{https://www.buildingsmart.org/standards/bsi-standards/industry-foundation-classes/}} for BIM models.

However, even as open standards are highly desirable, there is significant debate surrounding the CityGML standard.
As examples, \citet{Ledoux19} and \citet{20azul} report some of the issues of the CityGML standard as found by the developers point of view. 

Meanwhile, various users on the web also reported related issues, even if in a less academic context\footnote{Some examples from social media include:

\url{https://twitter.com/jamesmfee/status/748270105319006208}

\url{https://twitter.com/hugoledoux/status/1226065325327822848}

\url{https://twitter.com/MikeTreglia/status/866114277261930496}

\url{https://twitter.com/OSMBuildings/status/582884586134343680}

\url{https://twitter.com/GmLfun/status/1093512032350097409}

\url{https://twitter.com/GmLfun/status/1109117945240788992}

\url{https://stackoverflow.com/search?q=citygml}
}. The overall narrative is that the great number of ways in which the same models can be represented increases the implementation effort for software developers and reduces the quality of their implementations, therefore reducing interoperability and disappointing the expectations of end-users.

Nevertheless, these issues are mostly discussed informally by practitioners and academics and have not been tested systematically.
In order to gain more insight into the topic, the support of CityGML in software was investigated as part of the GeoBIM benchmark project\footnote{\url{https://3d.bk.tudelft.nl/projects/geobim-benchmark/}} (see Section~\ref{sec:intro2}) and reported in this paper.
Within the project, the approach to the study of the support for the two standards involved in the GeoBIM integration (IFC and CityGML) was conceived in parallel, also with the aim of understanding if one of the two offered more effective solutions that could be possibly borrowed by the other one in future developments.
However, the final outcomes of the two different tasks are very specific for each standard and deserve to be presented and discussed separately, considering the specificities of each case.
For these reasons, this paper, focusing on the results about Task 3, the benchmark task related to the support for CityGML, is written in tandem with \citet{noardo2020referenceI}, which describes Task 1 covering the support for IFC\@.
In order to allow each paper to be read on its own, the two papers share some information (\ie\ the first part of Section~\ref{sec:intro2}, explaining the general context and motivation of the study; Section~\ref{sec:metbench} covering the initial part of the methodology about the entire GeoBIM benchmark set-up, and Section~\ref{sec:mettask13} concerning some similarities in the methodology).
One further paper explores the parts of the project more directly related to the subject of integration, namely, conversion procedures and useful tools to georeference IFC models \citep{noardo2020tools}.

\section{The GeoBIM needs and the concept of this study
}\label{sec:intro2}

Two kinds of 3D information systems were increasingly developed, studied and used in recent times, revealing their potential in the related fields:

\begin{itemize}
	\item \textbf{3D city models}, which are used to represent city objects in three dimensions and advance previous 2D maps and other cartographic products, in order to support city analysis and management, city planning, navigation, and so on (\eg\ 
	\citet{Biljecki2015a,Kumar17,egusquiza2018energy,Jakubiec13,Liang14,Bartie10,Peters15,Nguyen12});
	\item	\textbf{Building Information Models (BIM)}, which are used in the architecture, engineering and construction fields (AEC) to design and manage buildings, infrastructure and other construction works, and which also have features useful to project and asset management (\eg\ 
	\citet{petri2017optimizing,haddock2018building,azhar2011building}%
).
\end{itemize}

Several international standards exist to rule the representation of the built environment in a shared way, to foster interoperability and cross border analysis and, consequently, actions, or to reuse tools, analysis methods and data themselves for research and possibly government.
Some examples of international standards are: the European Directive aiming at an Infrastructure for spatial information in Europe (INSPIRE)\footnote{\url{https://inspire.ec.europa.eu}}, aimed at the representation of cross border pieces of land in Europe, for common environmental analysis; the Land and Infrastructure (LandInfra)\footnote{\url{https://www.ogc.org/standards/landinfra}}, by the Open Geospatial Consortium (OGC), aimed at land and civil engineering infrastructure facilities representation; and the green building data model (gbXML)\footnote{\url{https://www.gbxml.org}}, aimed at the representation of buildings for energy analysis.

Nonetheless, the two dominant reference open standards for those two models are CityGML\footnote{\url{citygmlwiki.org}}, by the OGC, focusing on urban-scale representation of the built environment, and the Industry Foundation Classes~\citep{ISO16739:2013}\footnote{\url{ https://technical.buildingsmart.org/standards/ifc/}}, by buildingSMART, aimed at the very detailed representation of built works for design and construction objectives, first, but also intended to enable project management throughout the process, and asset and facility management in a following phase. Those standards are both intended to be very comprehensive and are therefore very wide and articulated.
They both use complex data models allowing for a wide variety of models using object-oriented representations, even if that comes at a cost of slower and more inconsistent implementations.

Due to the overlapping interests in both fields (meeting in the building-level representation), increasing attention is being paid to 3D city model-BIM integration (GeoBIM), where the exchange of information between geospatial (3D city models) and BIM sources enables the reciprocal enrichment of the two kinds of information with advantages for both fields, \eg\ automatic updates of 3D city models with high-level-of-detail features, automatic representation of BIM in their context, automated tests of the design, and so on~\citep{liu2017state,fosu2015integration,aleksandrov2019systems,kumar2019landinfra,niu2019logistics,Noardo19b,Arroyo-Ohori18a,kang2015study,Stouffs:2018kg,Lim:2019vh,sun2019evaluating}.

The GeoBIM subject can be divided into several sub-issues.

\begin{enumerate}
	\item First, the harmonization of data themselves, which have to concretely fit together, with similar (or harmonizable) features (\eg\ accuracy, kind of geometry, amount of detail, kind of semantics, georeferencing).
	
	\item Second, the interoperability is a fundamental key in the integration.
	It is important to note here, that before enabling the interoperability among different formats (\eg\ GIS formats and BIM formats), which is the object of the theme of point three below, the interoperability GIS-to-GIS and BIM-to-BIM itself is essential. That means that the formats of data have to be understood and correctly interpreted uniquely by both any person and any supporting software. Moreover, an interoperable dataset is supposed to remain altogether unchanged when going through a potentially infinite number of imports and exports by software tools, possibly converting it to their specific native formats and exporting it back. For this, it is desirable to rely on open standards.
	
	\item Third, the effective conversion among different formats, \ie\ transforming one dataset in a (standardised) format to another one in compliance with the end format specifications and features.
	
	\item Fourth, the procedures employing 3D city models and the ones based on BIM should be changed in order to obtain better advantages by the use of both, integrated, since those systems enable processes which are usually more complex than just the simple representations.
\end{enumerate}

The many challenges implied by the points above are still far from being solved, and one of the essential initial steps is actually to outline such challenges more sharply.

In particular, the second point (interoperability and involved standards) is often considered to be solved by standardization organizations.
It is indeed desirable to rely on open standards for this, because of the well-documented specifications of open standards enable longer-term support, as well as their genericity with respect to different software vendors, as opposed to closed point-to-point solutions that merely connect one proprietary system to another (and might be discontinued or stop working at any moment).
However, our previous experiences suggest that, unfortunately, the support for open standards in software is often lacking.

The researchers promoting this study (as users of data, advocates of open standards and developers of tools adopting such standards) have noticed, over their research and work activities, how the use of those standards in data and their implementations in software were not always straightforward and not completely consistent with the standard specifications either.
Many tools, when managing standardized data, do not adequately support features or functionalities as they do when the data is held in the native formats of the software.
In addition, software tools have limitations with respect to the potential representation (geometry, semantics, georeferencing) of data structured following these standards, or can generate errors and erroneous representations by misinterpreting them.

The standards themselves are partly at fault here, since they often leave some details undefined, with a high degree of freedom and various possible interpretations.
They allow high complexity in the organization and storage of the objects, which does not work effectively towards universal understanding, unique implementations and consistent modelling of data.
This is probably due to the fact that such standards often originate as amalgamations of existing mechanisms and compromises between the various stakeholders involved.
These experiences have been informally confirmed through exchanges within the scientific community and especially with the world of practitioners, who are supposed to work with (and have the most to gain from) those standardized data models and formats.
However, more formal evidence on the state of implementation of these open standards and what problems could be connected to the standard themselves have not been compiled so far.

For this reason, the GeoBIM benchmark project\footnote{\url{https://3d.bk.tudelft.nl/projects/geobim-benchmark/}}\(^,\)\footnote{\url{https://www.isprs.org/society/si/SI-2019/TC4-Noardo\_et\_al\_WG-IV-2-final\_report.pdf}} was proposed and funded in 2019 by the International Society for Photogrammetry and Remote Sensing (ISPRS)\footnote{\url{https://www.isprs.org}} and the European Association for Spatial Data Research (EuroSDR)\footnote{\url{http://www.eurosdr.net}}.
The aim of the benchmark was to get a better picture of the state of software support for the two open standards (IFC and CityGML) and the conversions between them, in order to formulate recommendations for further development of the standards and the software that implements them.
In addition, we tested two known major technical issues related to GeoBIM integration and which are known to be solved only partially in practice: the ability of tools and methods to georeference IFC and the conversion procedures between IFC and CityGML\@.

The relevant outcomes regarding the OGC CityGML standard are the subject of this paper.

\subsection{OGC CityGML: overview and knotty points}\label{sec:CityGML}

CityGML\footnote{\url{citygmlwiki.org} and \url{citygml.org}} (by Open Geospatial Consortium)~\citep{OGC2012a} is the most internationally widespread standard to store and exchange 3D city models with semantics in the geospatial domain.
It establishes a structured way to describe the geometry and semantics of city objects.

CityGML 2.0 (current version, considered for this project) contains classes structured into 12 modules, each of them extending the core module, containing the most general classes in the data model, with city object-specific classifications, \eg\ Building, Bridge, WaterBody, CityFurniture, LandUse, Relief, Transportation, Tunnel, Vegetation.
These modules contain one or more classes representing specific types of objects, which  differ in the way they are structured into smaller parts and the attributes that are expected for each.
The most developed and most used module in practice is the Building module.

Moreover, CityGML supports the possibility to further extend the schema through a standardized Application Domain Extension (ADE) mechanism~\citep{Biljecki2018}.
Some existing official ADEs, which could be useful for future tests of this project, are, for example, the Noise ADE, the Energy ADE, and the Utility network ADE\@.
However, it is known that ADEs have poorer software support, since most implementations would need to specifically encode the new objects and attributes added by an ADE\@.

CityGML proposed a very attractive management of useful concepts for user communities.
For example, it covers the most basic 3D city information with meaningful object-oriented representation, with deep hierarchies and complex relationships, which would be a more faithful representation of reality than a simpler relational database one.
However, the downside of this, is that sometimes complex and unusual connections of information to internal/external sources are used in some cases (\eg\ the prevalence of xlink-connected geometries within a file, the complex set of attributes used to store addresses, or the possibility to use references to external files through URIs), which could be a problem for the implementation of such model.

CityGML geometries are essentially the same for most classes: objects are represented as boundary surfaces embedded in 3D and consist of triangular or polygonal faces, possibly with holes.

CityGML as a data format is implemented as an application schema for the Geography Markup Language (GML) (CityGML uses version 3.1.1 of GML)~\citep{ogcgml31}.
It is an open format and human readable, that means that potentially, the information could be retrieved even if losing backwards compatibility in software.
However, GML presents many issues from a software developer point of view, since, for example, too many alternatives\footnote{\url{http://erouault.blogspot.com/2014/04/gml-madness.html}} are allowed even for simple objects, and a supporting application is supposed to foresee all possible combinations of them. 
The result of this complexity is that few software programs completely support all possible combinations, and most of its richness and power is lost (as the results of this research further demonstrate). 
An additional consequence of the kind of storage of such models is about their computational requirements: usually very large and complex files are produced, and it can be time- and resource- intensive to manage them properly in software.

As a possible solution to those issues, CityJSON\footnote{\url{https://www.cityjson.org}} (version 1.0.0)~\citep{Ledoux19} was recently released, providing a JSON encoding for a subset of the CityGML 2.0.0 data model. 
CityJSON follows the philosophy of another (non-standardised but working) encoding of CityGML\@: 3DCityDB~\citep{Yao18}.
That is, to store the models efficiently and allow practitioners to access features and their geometries easily.
The deep hierarchies of the CityGML data model are replaced by a simpler representation. 
Furthermore, some more restrictions are applied and one and only one way is allowed to represent the semantics and the geometries of a specific feature.
CityJSON is in the process to become an OGC community standard.
There is already a broad consensus around it, by users who are choosing it as an alternative to CityGML and many tools already developed to effectively work with it.
Some of the tests performed within this study use CityJSON as a pathway to process CityGML in order to manage the data effectively within software.

In this paper, the framework of the GeoBIM benchmark methodology is described, with a specific focus on the part of the project regarding the investigation of the support for CityGML and related results.

\section{Methodology}

\subsection{The GeoBIM benchmark general set-up
}\label{sec:metbench}

The benchmark was intended as a way to combine the expertise of many people with different skills, coming from several fields and interests, in order to describe the present ability of current software tools to use (\ie\ read, visualize, import, manage, analyse, export) CityGML and IFC models and to understand their performance while doing so, both in terms of information management functionalities, and possible information loss.
Moreover, since the big dimension of such standardised datasets often generate difficulties in their computational management, the ability to handle large datasets was a further part of the tests.

In particular, the four topics investigated in the benchmark are:

\begin{description}
	\item[Task 1] What is the support for IFC within BIM (and other) software?
	\item[Task 2]	What options for geo-referencing BIM data are available?
	\item[Task 3]	\textit{What is the support for CityGML within GIS (and other) tools?\footnote{This Task is the object of this paper.}}
	\item[Task 4]	What options for conversion (software and procedural) (both IFC to CityGML and CityGML to IFC) are available?
\end{description}

For this purpose, a set of representative IFC and CityGML datasets were provided~\citep{Noardo19a} and used by external, voluntary, participants in the software they would like to test in order to check the support in it for the considered open standard~\citep{Noardo3DGeoInfo2019a}.

Full details about the tested software and a full list of participants can be found in the respective pages of the benchmark website\footnote{\url{https://3d.bk.tudelft.nl/projects/geobim-benchmark/software.html} for the tested software and \url{https://3d.bk.tudelft.nl/projects/geobim-benchmark/participants.html} for the list of participants.}. The significant number of participants, balance in skills, fields of work, levels of confidence about the tested software (asked them to be declared) offered the possibility to limit the bias in the results.

The participants described the behaviour of the tested tools following detailed instructions and delivered the results in a common template with specific questions, provided as online forms.
In the end, they delivered both their observations and the model as re-exported back to the original standardised format (CityGML or IFC).

In order to cover the widest part of the list of software potentially supporting the investigated standards, we completed the testing ourselves, by searching the online documentation of both the standards and the potential software.

In the final phase of the project, the team promoting the study analysed the participants' observations, descriptions and delivered further documentation (screenshots, log files, related documents and web pages).
From this review, an assessment of the performances and functionalities of the tested tools was derived.
Moreover, the delivered models were validated and analysed using available tools, when possible, and/or through manual inspection (Section~\ref{sec:mettask13}).
This approach allowed the inquiry about the level of interoperability given by the standard and its software implementation, by comparing the results of the export with the imported model features.

It is important to notice that the test results are not intended to substitute the official documentation of each software.
Moreover, there were no expertise nor skill requirements to participate in the benchmark tests. Therefore, some information could be wrong or inaccurate, due to little experience with the tested software or the managed topics.
The declared level of expertise is intended to lower this possible bias.
Moreover, the benchmark team and the authors tried to double check the answers (at least the most unexpected ones) as much as possible, but the answers reported in the data were generally not changed from the original ones.
The eventual discrepancies between the best potential software performances and what was tested could anyway be showing a low level of user-friendliness of tools (and thus a degree of difficulty in achieving the correct result).

\subsection{The provided CityGML datasets}
A number of datasets from different sources were identified, pre-processed and validated for this benchmark activity (see \citet{Noardo19a} for details). The datasets were chosen to test both the most common features of such data and the main detected issues regarding the interesting but tricky aspects of the format.

Therefore, a large LoD1 file was chosen to test the support for quite simple but extended dataset: \ie\ the whole city of Amsterdam in LoD1, covering the representation of many city-related objects and useful to test the software and hardware related performances. Second, a two-LoD file (LoD1 and LoD2) representing a district in Rotterdam was aimed at testing the support of different LoDs stored in the same file; and finally, the more complex geometries (and related different semantics) needed for an LoD3 representation were tested by means of the synthetically generated file \textit{BuildingsLoD3.gml} (\Reftab{tab:CityGMLdata}).
LoD4 models were not included in the provided datasets, since nearly no examples can be found in practice.

 \begin{table}[H]
 	\centering
 	\small
 	\begin{tabular}{|m{1.7cm}|m{5cm}|m{2cm}|m{3cm}|m{4cm}|}
 		\hline
 		\textbf{Name} & \textbf{Description} & \textbf{Di\-men\-sion} & \textbf{Source} & \textbf{Aim}  \\ \hline
 		\textit{Amster\-dam.gml} & Seamless city model covering the whole city of Amsterdam, including several CityGML city entities (vegetation, roads, water, buildings, and so on). Level of Detail (LoD) 1. & 4.06 GB                                         & Generated through 3dfier by TUDelft\footnotemark & Test of the hardware-and-software connected performances (it is a very heavy model), and support for the included city classes. \\ \hline
 		\textit{Rotterdam\-LoD12.gml}                  & Textured CityGML model of one district in Rotterdam, including only Buildings in LoDs 1 and 2.                                                                            & 33.91MB/ 154.4MB (with textures) & Municipality of Rotterdam (NL)                                                                                 & Test of the support for multiple LoDs and textured files.                                                                \\ \hline
 		\textit{Buildings\-LoD3.gml}                   & Procedurally modelled buildings in LoD 3 through                                                                                                                            & 1.33 MB                                         & Generated through Random3Dcity \citep{Biljecki:2016td}.\footnotemark & Test of the support for LoD 3 files and related classes.                                                         \\ \hline
 	\end{tabular}
	\caption{Provided CityGML data for the GeoBIM benchmark 2019}%
\label{tab:CityGMLdata}
 \end{table}
\footnotetext[17]{\url{https://github.com/tudelft3d/3dfier}}
\footnotetext{\url{https://github.com/tudelft3d/Random3Dcity}}

\subsection{Answers analysis about the support for CityGML
}\label{sec:mettask13}

The methodology for analysing the results about the support of software for IFC (Task 1) and CityGML (Task 3) are very similar, since they were also conceived to test similar issues concerning interoperability and the ability of software to keep files consistent with themselves after import-export phases.

The initial part of results analysis (Section~\ref{sec:swsupptask3}) is qualitative, providing the description of software support and functionality based on the delivered answers.

The complete answers and documents delivered in the online templates \citep{noardofrancesca2020-reportsTask3}\footnote{\url{http://doi.org/10.5281/zenodo.3966987}} were double checked for correctness and consistency with respect to the asked questions. 
However, due to the nature of the initiative, we trusted the delivered information about the software, double checking it with new tests only in cases of inconsistent answers in different tests about the same software, or eventually, unexpected answers.
In these cases, we also considered the level of expertise of the participant to assess if further checks were actually needed.

The delivered answers in the templates were critically assessed, cross-checking them with the different tests about the same software and the attached screenshots. A score about each aspect considered for the assessment of general support and software functionalities is assigned, as: 1-full support; 0.5-partial support; 0- no support. Those are synthesized in a table (\Reftab{fig:task3synthesis}), from where it is also easier to deduce possible patterns across many issues for a single software package or across many software packages for a single issue.

The definition of software groups are getting increasingly fuzzy, since the functionalities of all of them are continuously being extended and now tend to overlap with each other.
However, in the tables, and more generally, in the analysis, in order to help the detection of possible patterns, the tested software are classified considering the criteria that usually guide the choices made by users, based on their different needs for specific tasks:

\begin{itemize}
	\item \textit{GIS} is expected to combine different kind of geodata and layers and make analysis on them, structured in a database, in a holistic system;
	\item \textit{`Extended' 3D viewers} are likely software that were originally developed for visualising the 3D semantic models, including georeferencing, and query them. They were (sometimes later) extended with new functions for applying symbology or making simple analysis.
	\item \textit{Extract Transform and Load (ETL)} software, and conversion software, are expected to apply some defined transformations or computations to data;
	\item \textit{3D modelling tools} have good support for geometry editing, but is not originally intended to manage georeferenced data nor semantics;
	\item \textit{Analysis software} are intended specifically for few kinds of very specific analysis (\eg\ energy analysis);
	\item \textit{BIM} software, are intended to design buildings or infrastructures according to the the BIM methods.
\end{itemize}

The investigated issues, reflected in the different sections of the provided templates, regarded mainly the support of the software for the two standards (how the software read and visualise the datasets) and the functionalities allowed by the software with standardised datasets (what is it possible to do with such data).
In particular, the test about the support was intended to test: how is the georeferencing information in the files read and managed; how are the semantics read, interpreted and kept after the import; and how is the geometry after the import.
Georeferencing is about the ability of the software to locate and visualize the data at the correct georeferenced coordinates, together with recognizing the correct coordinate reference system (CRS), as read within the file.
Semantics are the thematic data associated to the objects of the model, which are structured in hierarchies of classes, reciprocally related by means of the  relationships defined in the data model.
In this case, the reference data model is ruled by the CityGML standard.
Further information is associated to such entities as attributes.
The third point is about geometry, \ie\ the way each object is modelled spatially and how such geometry is stored (\eg\ as a solid, as a surface and so on).

Moreover, some additional questions for Task 3 were intended to investigate if any additional step or conversion was necessary to use the file within the tested software, or straightforward workflow was sufficient (\eg\ opening the software and importing the file by pushing a button).

Additional questions therefore are: what kind of formats can be managed (is a file supported through the GML encoding, or a conversion is eventually needed in addition or alternatively, to the CityJSON format or to a database, mainly SQL)?
Does the software support the file out of the box, or is some specific plug-in or add-on needed?

\begin{itemize}	
	\item What kind of visualization is enabled (3D, 2D, with textures, with specific themes);
	\item What kind of editing is possible (attributes, geometry, georeferencing);
	\item What kind of query (query the single object to read the attributes, selection by conditions on attributes, spatial query, computation of new attributes);
	\item What analysis are allowed. This topic is more complex, since very different analysis can be possible. Therefore we summarized it by defining two analysis types: `Type 1' is any kind of analysis regarding the model itself (like geometric or semantic validation), and `Type 2' are the simulations and analysis about the performances of the represented object (\eg\ a building) with respect to external factors, in the city or environment (\eg\ shadow, noise, energy, etc.).
	\item Final issue: Is it possible to export back to CityGML\@?
\end{itemize}

One more aspect that was asked to the testers of Task 3 (support for CityGML), and checked in the delivered results, was the support for ADEs, although it was just checked in theory, since no ADE datasets were provided.

Moreover, the support for each of the delivered datasets was noted, given the specific features as follows:
multi-LoD management (through \textit{RotterdamLOD12.gml} dataset), LoD3 management (through \textit{BuildingsLOD3.gml}) and a large LoD1 model (\ie\ \textit{amsterdam.gml} dataset).

This first parts provide a reference about the tools themselves for people intending to use standardised information.
In addition, the most challenging tasks and most frequent issues for the management of standards were supposed to be pointed out.

A second, more quantitative, part of the analysis considers the delivered models exported back to CityGML \citep{noardofrancesca2020-exportsTask3}\footnote{\url{https://doi.org/10.5281/zenodo.3966915}} from the tested software (Section~\ref{sec:interoptask3}). 
The numbers and types of features of such files were calculated and compared to the same features in the initial datasets that were provided for the test.

The semantics were checked, in terms of number of entities and relationships, as computed by the statistics tool related to the KIT FZK viewer\footnote{https://www.iai.kit.edu/1302.php}. Moreover, the presence and consistency of attributes was also checked by means of manual inspection in 3D viewers (FZK and azul\footnote{https://github.com/tudelft3d/azul}).

In addition, the CityGML schemas were validated by means of the GML schema validator related to the FZK viewer and the CityGML schema validator\footnote{\url{http://geovalidation.bk.tudelft.nl/schemacitygml}}.

The number and kind of geometries were also counted by the FZK statistics tool, further supported by manual inspection, in some cases.
The val3dity\footnote{\url{http://geovalidation.bk.tudelft.nl/val3dity}} validation tool~\citep{ledoux_validation_2013} allowed, finally, the test of the validity of re-exported geometries.

This allowed us to assess the level of interoperability that the connected standards-tools can actually reach in the different cases: \ie\ can the data be imported and re-exported without any change?

A further assessment (Section~\ref{sec:swperftask3}) was intended to evaluate the software and hardware connected performance.
The times declared by the testers were compared for the three datasets to see if their computational weight could affect their management within software.

Given the complexity of measuring software performance to the closest second, this was not requested from the users. Instead, they were asked to provide an approximate timing value for each test, according to a classification that was proposed following the way they could affect the perception or the work of a user, as explained in the following list:

\begin{itemize}
 \item It is almost immediate (good!)
 \item Less than a minute (ok, I will wait)
 \item 1--5 minutes (I can wait, if it is not urgent)
 \item 5--20 minutes (in the meantime I do other things)
 \item 20 minutes--1 hour (I cannot rely on it for frequent tasks)
 \item more than 1 hour (I launch my process and go home, definitely ineffective for regular work)
\end{itemize}

Other options included reporting if the software crashed or if the task was not possible with the software provided, and participants were also asked to provide information about the specification of the machine, as this may impact overall performance of the software.  
Due to their diverse levels of size and complexity, timing results are summarised for the individual datasets.

\section{Results: support of software for CityGML}\label{sec:resultstask3}


\subsection{Tested software against support for CityGML}\label{sec:swtask3}

In order to ensure the coverage of most of the software solutions to manage CityGML, we checked that the main currently used GIS software packages were tested, and in addition we tried to ensure to have the tools declaring some support for CityGML tested as well.

The tested software packages were 15 in total, with several tests for some of them, likely covering a most of the possible solutions. The full list is reported in Table~\ref{tab:swtask3tab}.
In the table, they are organised based on the kind of software and divided into: open source, proprietary and freeware (but not open source) software.
Moreover, the levels of expertise of participants making the tests (from L1 the least to L4 the most expert) are also reported.

\begin{table}[H]
\centering
\small
	\begin{tabular}{|p{3cm}|p{3cm}|p{5cm}|p{4cm}|}
		\hline
		& \textbf{Open Source} & \textbf{Proprietary} & \textbf{Freeware} \\ \hline
		\textbf{GIS software} & QGIS [L1+L2+L3] & ESRI ArcGIS [L1+L2]

		ESRI ArcGIS Pro [L2]& \\ \hline
		\textbf{`Extended' 3D viewers} & 
		& Safe software FME data inspector [L3]

		eveBIM [L4]

		M.O.S.S. novaFACTORY+WEGA-3D [L3]

		1Spatial Elyx 3D [L2] & FZK Viewer [L1+L2] 

		Hexagon AB Tridicon CityDiscoverer Light [L1] \\ \hline
		\textbf{ETL and conversion software} & 3DcityDB[L3] & Safe software FME [L3] & \\ \hline
		\textbf{3D modeling software} & Blender+CityJSON plug-in [L1] & ESRI CityEngine [L1] & \\ \hline
		\textbf{Analysis software} & & Kaemco CitySim Pro [L1] & \\ \hline
		\textbf{BIM software} & & Autodesk Infraworks [L1] & \\ \hline
	\end{tabular}%
\caption{Tested software for Task 3 --- Support for CityGML within software.}%
\label{tab:swtask3tab}
\end{table}

Some of the software packages were tested several times, especially the most known and used GIS tools, where the inclusion of 3D city model data could be the most interesting and useful, besides natural, since they are the current tools employed to manage geoinformation.

\textit{QGIS}\footnote{\url{https://www.qgis.org/en/site/}} was tested 4 times, plus one partial answer, by participants having different levels of expertise (beginners, current users and experts). The four tests conducted adopted different approaches, though: built-in support for GML, import of the file after conversion to CityJSON and the use of the CityJSON loader plug-in.
Additional methods are the import through \textit{citygml4j}, and the use of \textit{GMLAS}\footnote{\url{https://gdal.org/drivers/vector/gmlas.html}}.
However, the \textit{GMLAS} procedure was not successful for the import of the data.

The CityJSON\footnote{\url{https://www.cityjson.org/software/}} implementation was used to enable the import of the datasets in Blender, by means of the CityJSON plug-in\footnote{\url{https://github.com/cityjson/Blender-CityJSON-Plugin}}, which could be useful to model and edit the models.

\textit{ESRI ArcGIS} was also tested multiple times, both in the standard version (3 tests) and in the Pro version (2 tests). Other popular software were the \textit{FME} (Feature Manipulation Engine)\footnote{\url{https://www.safe.com}} from \textit{Safe software}, tested twice, and the freeware extended 3D Viewer KIT \textit{FZK Viewer}\footnote{\url{https://www.iai.kit.edu/1648.php}}, also tested twice (by testers having levels of expertise 1 and 2).

Moreover, other applications were considered (raising the number of tested software to 26), selected on the base of information found on the CityGML Wiki under `software'\footnote{\url{http://www.citygmlwiki.org/index.php/Commercial\_Software}}\(^,\)\footnote{\url{http://www.citygmlwiki.org/index.php?title=Freeware}}\(^,\)\footnote{\url{http://www.citygmlwiki.org/index.php?title=Open\_Source}} or in the websites of software programs declaring support for CityGML\@. However, some of them, the ones for which the full test was not performed, did not actually support the data or were no more available, namely:

\begin{itemize}
	\item \textit{Autodesk Landxplorer CityGML viewer} is quite outdated, it is possible to find information about it until 2011 approximately\footnote{\url{http://download.autodesk.com/us/landxplorer/docs/LDX11_Studio/index.html?topic.htm}} and no download was found except for an external website publishing an executable for its 2009 version\footnote{\url{https://www.cadforum.cz/cadforum\_en/download.asp?fileID=1064}};
	
	\item \textit{Bentley Map} and \textit{Bentley Microstation} were tested, but it was not possible to import CityGML datasets;
	
	\item \textit{SketchUP+CityGML plugin}\footnote{\url{https://forums.sketchup.com/t/citygml-plugins-for-sketchup/24921}} is outdated: it worked for \textit{SketchUP} v.2015, whilst the newest one is v.2019;
	
	\item \textit{Cesium viewer} and \textit{Google Earth} work only through conversion. There are different tools available for that purpose but the participants in this study chose the \textit{3DCityDB} converter. \textit{3DCityDB} loads CityGML into relational tables in \textit{PostGIS} or \textit{Oracle Spatial} and can export into CityGML, KML/COLLADA and glTF formats. Rendering is very efficient with these formats at the expense of loosing complex semantics and relationships.
	
	\item \textit{Sidefx Houdini} was proposed for testing by one of the participants, but it does not work with CityGML\@;
	
	\item \textit{iTOWNS}\footnote{\url{http://www.itowns-project.org}} does not support CityGML, unless, probably through some conversion to \textit{PostGIS}\footnote{\url{https://github.com/Oslandia/workshop-3d-itowns/blob/master/0\_Data/1\_Buildings.md}};
	
	\item \textit{HALE studio}\footnote{\url{https://www.wetransform.to/products/halestudio/}} should also be able to support CityGML, but the function to import CityGML was not actually found in the software.
\end{itemize}

\subsection{Software support for CityGML}\label{sec:swsupptask3}

In this section, the qualitative analysis of the delivered answers of participants \citep{noardofrancesca2020-reportsTask3}\footnote{\url{http://doi.org/10.5281/zenodo.3966987}} describing the software tools and the tests is reported.
First the observations about how the GML format is read by software are summarized in Section~\ref{sec:load}.
A second level of the test was about checking that the data was interpreted and read correctly, that means their georeferencing information (Section~\ref{sec:loadgeoref}), their semantics (Section~\ref{sec:loadsem}) and their geometry (Section~\ref{sec:loadgeom}) were not lost in the conversion to the native formats read by software but remained actually interoperable instead.
Those answers are summarized in \Reftab{fig:task3synthesis}.

\begin{table}[htbp]
	\centering
	\includegraphics[width=1\linewidth]{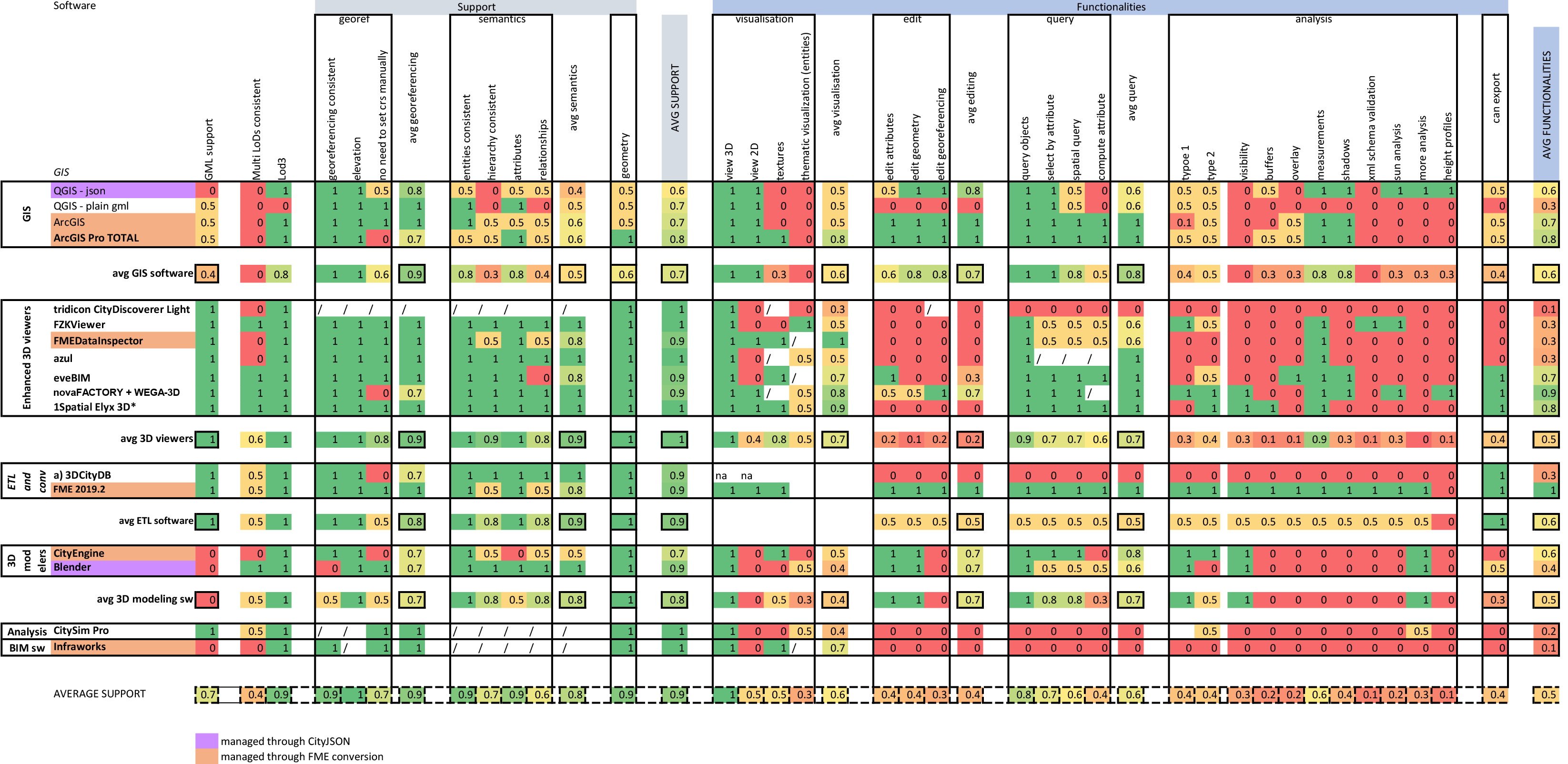}
	\caption{Synthesis table of the delivered tests regarding support for CityGML (benchmark Task 3). The colour scale is assigned according to the scores from 1-full support (green) to 0-no support (red). In the table, the software supporting CityGML through conversions to other formats, namely CityJSON or through \textit{FME}-based scripts are indicated in purple and orange, respectively.}
	\label{fig:task3synthesis}
\end{table}

\subsubsection{Loading of GML data}\label{sec:load}

First, the support for loading (City)GML files directly by the software was evaluated.
This functionality would allow conversions between different formats to be avoided, ensuring easier use of such data by experts of other domains.
Also, it would prevent the introduction of further errors and inconsistencies to the data, due to the conversion process.

In the case of GIS software (\textit{ArcGIS} and \textit{QGIS}), direct import of GML files was only correctly done with LoD1 data, which is basically 2.5D.
Moreover, very few functionalities are enabled in this case, being only possible to visualise and query the data, without editing. 
The other two datasets --- that is the LoD3 and multi-LoD files, which are both ``real'' 3D --- were interpreted in a completely wrong way: geometries were only loaded as points for the multi-LoD dataset; and only attributes with no geometries were loaded for the LoD3 dataset.

To import the data consistently, additional software or specific plug-ins were required in both cases.
For \textit{ArcGIS}, the `Data Interoperability' toolbox was used, which is based on the Safe Software Feature Manipulation Engine (FME).
For \textit{QGIS}, several plug-ins exist, out of which only one worked (CityJSON Loader\footnote{\url{https://github.com/tudelft3d/cityjson-qgis-plugin}}) which requires a conversion to the CityJSON format.
The conversion to a relational database through \textit{3DcityDB} would also work in both cases, but again that involves a conversion and external tools in the process.

Different is the case for the group of `Extended 3D viewers', since they were implemented specifically to work with the (City)GML format and can, therefore, read it directly.
The same is true for the ETL and conversion tools, able to interpret the GML format, since their aim is generally to transform it to something else.
Also in \textit{CitySim Pro} a specific import for CityGML was implemented, and works.

That is not the case of 3D modellers and BIM software, where external tools are needed: the proprietary software \textit{ESRI CityEngine} and \textit{Autodesk Infraworks} only manage their own proprietary format, therefore the import through a connected \textit{FME}-based converter is necessary.
\textit{Blender}, open source, is able to manage several more standardised formats, instead, but the specific plug-in is needed and, furthermore, this is able to import the data in CityJSON, that means the data should be converted to CityJSON\footnote{For the conversion citygml-tools can be used: \url{https://github.com/citygml4j/citygml-tools}}.

To summarise this issue, 70\% of the tested software can read GML directly, however, most of them (9 out of 15) were specifically programmed for CityGML\@.
The remaining ones need conversions that usually go through \textit{FME} processing (especially the proprietary ones) and/or through the conversion to CityJSON\footnote{An \textit{FME} reader/writer for CityJSON has now been implemented (\url{https://github.com/safesoftware/fme-CityJSON}), so that it likely could also be used for importing CityJSON data. However, this functionality was not yet available during the data collection phase of the benchmark.}, or require the use of plug-ins.

Another point regarding the support of software for the CityGML data model is about the more complex kinds of geometry management, once imported, namely: the consistent interpretation and visualization or use for analysis of multi-LoDs datasets and functioning with LoD3 data.

The consistent interpretation of multi-LoD is intended as the possibility to read the information of the CityGML objects associated to the various LoDs geometries at once, among which it is possible to choose from the same interface, for both visualisation or eventual use of that in analysis. This was tested through the \textit{RotterdamLoD12.gml} dataset.
Only 40\% of the tested software is able to manage them consistently:

\begin{itemize}
	\item No consistency in GIS (both read together and superimposed);
	\item Partial support from the ETL and conversion tools (in \textit{FME} it is either considered as a unique aggregate or it is possible to choose to upload them separately and in \textit{3DcityDB} it is necessary to choose which LoD to work with);
	\item Being based on \textit{FME}, the same is true for \textit{CityEngine} and \textit{Infraworks};
	\item Partial support is in \textit{CitySim Pro,} since only LoD2 and LoD3 are allowed, therefore it was difficult for us to test it;
	\item The issue is well managed by most of extended 3D viewers (except for \textit{tridicon CityDiscoverer Light}, \textit{FMEDataInspector} and \textit{azul});
	\item \textit{Blender} can also manage them consistently (through the CityJSON format).
\end{itemize}

LoD3, tested through the \textit{BuildingsLoD3.gml }dataset is well supported by all of the tested software, once the necessary conversions are made (it is not read only by \textit{QGIS} if the GML format is used).

\subsubsection{Loading georeferencing}\label{sec:loadgeoref}

Looking at georeferencing, the information about coordinate reference systems (CRS) can be consistently read by all the software managing the files (except for \textit{Blender}, which is not intended to be managing georeferenced objects), as well as the heights, in few cases it is necessary to set the CRS manually, for example in \textit{QGIS} or ArcGIS in some cases, in \textit{novaFACTORY} and it also has to be explicitly set in \textit{FME} and, consequently, \textit{CityEngine}. No relevant problems are found for this aspect, as expected by software born to manage geoinformation.

\subsubsection{Loading semantics}\label{sec:loadsem}

More difficult management is observed for semantics:

\begin{itemize}
	\item `Extended' 3D viewers as well as ETL and conversion tools can usually interpret the semantics properly, as consistency of entities, hierarchies, attributes and further relationships, with some exceptions, for example, the hierarchy and relationships are managed through parent-ids in relational database fashion in some cases (\textit{FME} and related software);
	\item \textit{Tridicon CityDiscoverer light }has no functionality to read neither georeferencing nor semantics information;
	\item In the GIS tools the semantics is converted to the internal, relational, structure of data, therefore some part of the information is always lost or managed through different tables, connected by means of IDs, that make the structure way more complex and less usable.
\end{itemize}

In \textit{QGIS} the testers\footnote{In \citep{noardofrancesca2020-reportsTask3}, `QGIS test 3'.} pointed out that sometimes the same entity information can be lost (wall, door, building), but the attributes result listed under a generic \textit{`cityobjectmember'}, when managed through the GML format.
In the CityJSON case, instead, the geometries are correctly recognized, but they cannot be accessed through the attribute table.
The entity name (Building, BuildingPart and so on) is kept in the attribute `type' of the CityJSON format.
The relationships are managed through ids stored in the attribute tables. 
The same management of attributes is sometimes made more complex by their storage in different tables, connected through IDs.


\subsubsection{Loading geometry}\label{sec:loadgeom}

The tested tools had little possibility to assess the geometry validity and correctness, and more results are found after analysing the exported models (Section~\ref{sec:interoptask3}).
However, the geometries look generally good, except in GIS software when directly importing GML files without conversion. In that case we already discussed that the 3D (LoD3) or multi-LoD geometries were not read correctly.

A secondary question about the support for CityGML regarded the possibility to manage Application Domain Extension (ADE) information. This is managed by \textit{FME} and some of the software using \textit{FME}-based procedures (\textit{ArcGIS}), \textit{FZK Viewer}, through the addition of the schemas in the software files, \textit{eveBIM}, \textit{novaFACTORY} and \textit{Blender}. In none of them it is, however, possible to use the information for analysis, except for \textit{FME} 
and \textit{ArcGIS}, as stated in the answers.
In the other cases such information can only be viewed and sometimes queried.

\subsubsection{Using CityGML data}\label{sec:swfunctask3}

The functionalities asked for testing were: visualisation, editing, query and analysis.

All the tested tools can visualise the data in 3D and some of them (the GIS software, the ETL and conversion tools, plus \textit{1spatial Elyx3D} and \textit{FME Data Inspector}) can also in 2D, which is not, however, the priority of those tools nor of those data.

Partial support also regards the visualisation of textures, consistently provided by half of the software tools: \textit{ArcGIS Pro}, \textit{FME Data Inspector}, \textit{FME}, \textit{CityEngine}, \textit{Infraworks} (all based on \textit{FME} conversion and enabled to visualise textures in their native formats), \textit{eveBIM} and \textit{1Spatial Elyx3D}.
The thematic visualisation (\ie\ the application of symbology based on variables such as the name of entities, attribute values, queries and so on) is enabled as association of different colours to different entities in \textit{azul}, \textit{novaFactory}, \textit{1Spatial Elyx3D}, \textit{Blender}, \textit{CitySim Pro}.
\textit{FZK Viewer} also allows some thematic symbology based on the values of some attributes.
A drawback of this is that only limited amount of pre-set thematization are allowed, without full customization possible.
But it is still the most advanced application for this functionality, which could be very useful for current GIS users to use 3D information.
\textit{QGIS} also allows the application of symbology to 3D data in the 3D viewer.

The possibility to edit the data is also important for users. In this task, the GIS tools offer better support. Usually, it is possible to edit attributes, geometry (for example, by moving vertexes) and georeferencing.
This is however not possible with GML data in \textit{QGIS}\@.
Limited editing is possible in enhanced 3D viewers: for example, it is only possible to remove objects from the \textit{FZK Viewer}, \textit{novaFACTORY} can edit attributes and geometry through a module called `Feature3D' and additional plug-ins, as well as \textit{1Spatial Elyx3D} can edit attributes only if the data are converted to relational DBMS and imported in that format.

We can notice that, unfortunately, the tools which can best read the GML format (the enhanced 3D viewers) are the least able to manage the editing.
This is likely because good functionality is achieved partly through a conversion to an optimised internal data model, from which it is hard to export back to CityGML\@. 

Moreover, any kind of change is possible through the workflows of \textit{FME}, and partially in \textit{CityEngine} and \textit{Blender}, where it is possible to edit attributes and geometry. The format edited in \textit{CityEngine} is the native software format as converted by \textit{FME}, while in \textit{Blender} the CityJSON format is managed and can be edited.
Finally, \textit{CitySim Pro} allows the editing of some energy-related parameters, external to the dataset, according to its scope. Also, \textit{3DcityDB} does not include editing, query and analysis functionalities, but of course those are not the aim of the software and it is not considered a limitation.

The query functionality is another important requirement for working with data effectively.
The synthesis table (\Reftab{fig:task3synthesis}) shows that most of the tested software, especially GIS and 3D viewers (except \textit{tridicon CityDiscoverer Light}), allow to query the objects directly (by `clicking' on them) to read the object attributes.
\textit{CityEngine} and \textit{Blender} also allow that. 
More complex queries, such as the selection of entities based on rules, are possible in some of the software.
They are well managed in GIS, although the 2D footprint of the geometry is often considered when running spatial queries.
In 3D viewers, except for \textit{tridicon CityDiscoverer Light}, all sort of queries are also possible, even if reduced in some cases, for example, very simple queries can be performed in \textit{FME Data Inspector}, like looking for one attribute in a table, and in the \textit{FZK Viewer} they are quite predefined.
In the case of \textit{FME} and \textit{Blender}, instead, any kind of query is supported but it has to be specifically programmed by the user: through the use of \textit{FME} transformers in \textit{FME} or in \textit{Python} in the case of \textit{Blender}.

The data analysis functionalities of the software (regarding geometry, semantics and georeferencing validity) and the simulation of interaction of the represented objects with their context (Type 1 and Type 2, respectively) were considered separately.
This topic will not be treated exhaustively, probably, since not all the participants reported on them with the best possible accuracy and, moreover, the least expert of them could not know how to get to the least apparent analysis tools. Nevertheless, from the tests it is possible to see that, generally, partial support to analysis is in GIS tools, that mainly work with 2D information. Sometimes additional plug-ins or toolbox can perform further 3D analysis.
Such tools are quite new and little developed even for the 3D information in the same software native format, therefore we could not have the chance to understand if they work differently when used on CityGML data.
In \textit{ArcGIS}, the \textit{3D Analyst extension} is necessary, that used to deal with 2.5D information like Digital Terrain and Surface Models, and probably it is being extended also for real 3D.
Moreover, it is possible to program more analysis through \textit{Python} scripts, which is however not so user-friendly for the least expert users.
Very little analysis is possible within 3D viewers, most of the times measurements are possible on the model, some energy analysis can be done in \textit{FZK Viewer} and \textit{novaFACTORY}, shadows in some cases (\textit{eveBIM} and \textit{novaFACTORY}), overlay and buffers and visibility analysis can be performed in \textit{novaFACTORY}, \textit{1Spatial Elyx 3D}, \textit{CityEngine} and \textit{Blender}.
Moreover, the validation of the schema can be done in \textit{FZK Viewer} and the same, plus some geometric validation and many other analysis are possible in \textit{FME}, which is the one most supporting this kind of functionalities, through building workflows by means of its transformers.
The specific goal of \textit{CitySim Pro} is to run energy analysis. However, in this test those did not work with the provided CityGML datasets.
The implementation of analysis tools in software supporting CityGML is therefore not very developed yet, with some exception for visibility and shadow analysis and some energy tools which is available sometimes.
The availability of tools analysing 3D information is generally to be enhanced.

Finally, few software packages can export 3D city models back to CityGML\@.
Notably, this is true even when the software is able to import it through \textit{FME} (\eg\ \textit{CityEngine}, \textit{Infraworks}), which itself has CityGML export functionality.
Some GIS software can export it through the same tools they used for the import (\textit{ArcGIS Data Interoperabilty toolbox} and \textit{CityGML-toolbox for CityJSON} in \textit{QGIS}). 
In some 3D viewers, the data are not converted at all, therefore they are just saved with the new eventual changes without being exported (that implies, instead, a conversion).
Moreover, the ETL and conversion tools are developed on purpose to run import-export processing, and therefore are able to export too.

\subsection{Software performances with CityGML data}\label{sec:swperftask3}

A total of 21 different reports were returned, for 15 different software packages.
In particular, multiple results for \textit{Esri ArcGIS Pro} (2 sets), \textit{Esri ArcGIS} (3 sets), \textit{QGIS} (3 sets) and \textit{FME} (2 sets).
These offer the opportunity for timing comparisons to investigate the impact of hardware on software performance. 

\begin{figure}[H]
	\centering
	\includegraphics[width=0.95\linewidth]{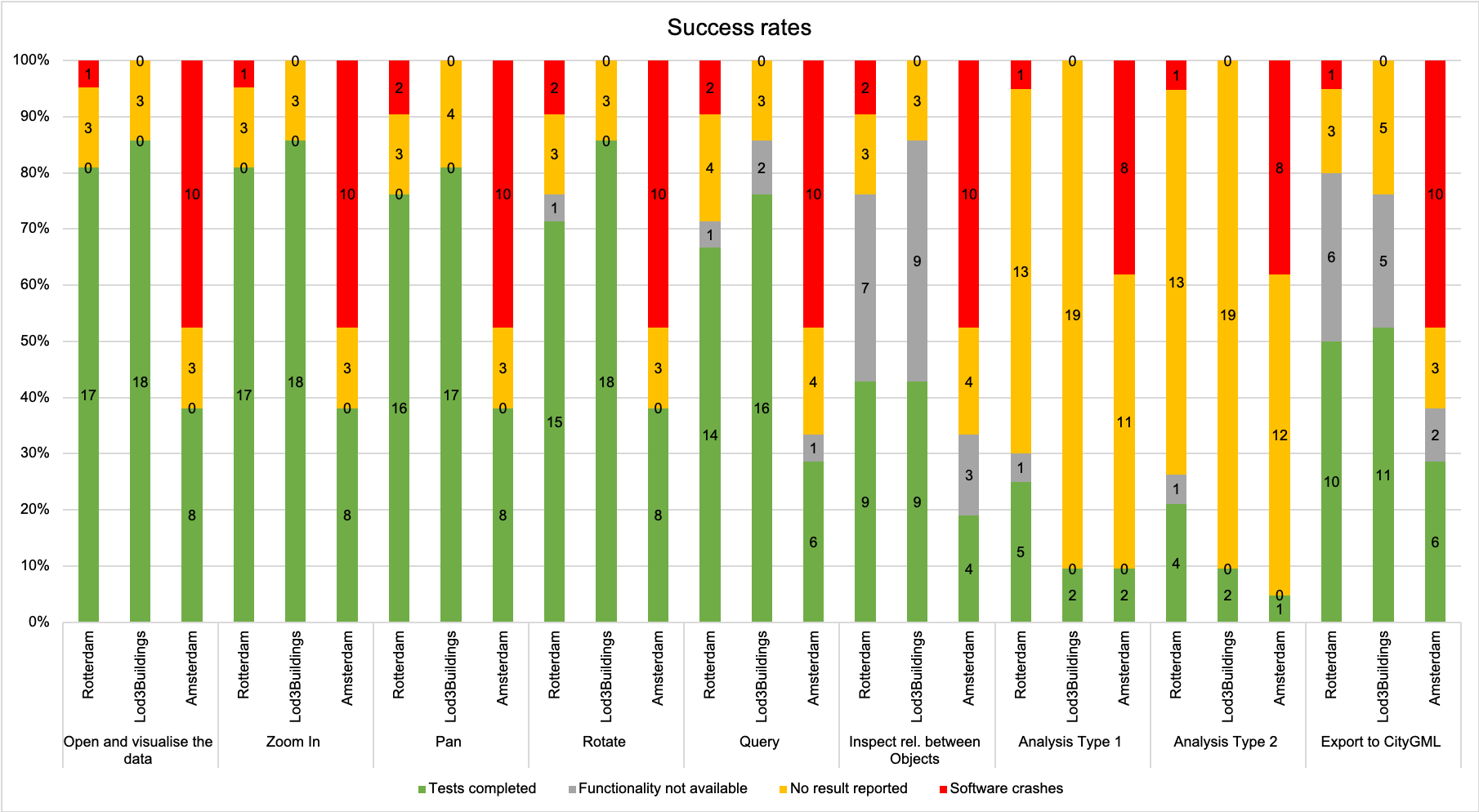}
	\caption{Graph reporting the number of tools associated to their success rate in the indicated tasks.}%
	\label{fig:successrates}
\end{figure}
\begin{figure}[H]
	\centering
	\includegraphics[width=0.95\linewidth]{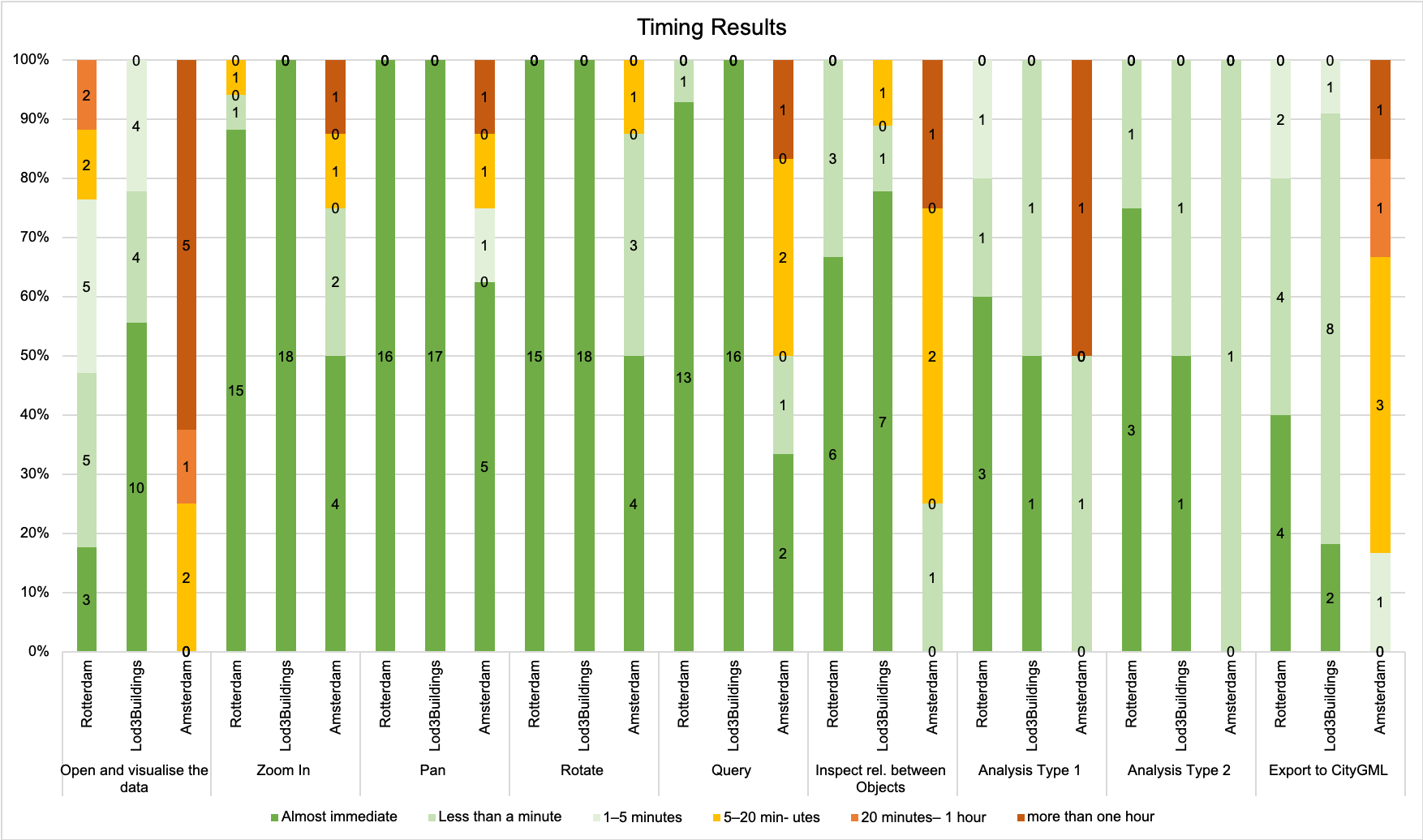}
	\caption{Graph reporting the number of tools associated to their timing results in the indicated tasks.}%
	\label{fig:timinggraph}
\end{figure}

\Reffig{fig:successrates} gives a summary of the success rates returned for the tests on the three datasets.
\Reffig{fig:timinggraph} gives the count of the different timing values for the successful tests.

Note that in some cases users reported results for some of the tests but did not report results for all of the tests (``No result reported'').
Additionally, some users typed in comments such as ``no error'' instead of giving specific timing.
These are included in the ``No result reported'' count.

For the \textit{RotterdamLoD12.gml} dataset, 2 out of 17 of the successfully completed tests took over 20 minutes to open the dataset (\textit{City Engine} and \textit{ArcGIS}, with the second \textit{ArcGIS} test reporting that this task took between 5 and 20 minutes).
The vast majority of the software packages tested could zoom and pan the data immediately with only one package taking between 5 and 20 minutes to zoom into the dataset (\textit{eveBIM}).
Only 10 out of the 21 tests report a successful export to CityGML, with all of these taking less than 5 minutes to execute.

With the \textit{BuildingsLoD3.gml} dataset, all 18 of the successful attempts to open the dataset took less than 5 minutes.
This reflects the small size of this dataset, and similarly all 11 successful attempts to re-export the data took less than 5 minutes.
Rotation, zoom and plan are very rapid --- but one report notes a time of between 5 and 20 minutes to inspect relationships between objects (\textit{Esri ArcGIS}).

The results with the \textit{amsterdam.gml} dataset demonstrate clearly the impact of the larger dataset on the tests carried out, with only 8 out of 21 reports indicating that the software was able to handle the data (48\%), with an additional 38\% reporting software crashes.
The three `no result reported' here relate to \textit{QGIS}, where the testers either displayed the data as points or first converted the data to CityJSON, which is not what the task required. 

None of testers reported a time of less than 5 minutes to open the dataset, and 5 reporting a time of over 1 hour.
\textit{3DCityDB} reported an export time of 1--5 minutes, perhaps due to its bespoke development for CityGML, in contrast to the generic functionality offered by many of the other software packages. 

\subsubsection{Multiple Tests on Same Software Packages}

The crowdsourcing approach taken in this project resulted in multiple participants testing the same software, providing the opportunity for comparison.
However, the three \textit{QGIS} tests were eliminated from the comparison as one imported the data into points and the other two created CityJSON data.

\begin{itemize}
\item Comparing the \textit{FME Data Inspector} tests, many of the results obtained were identical in terms of performance time.
A slight difference (1--5 minutes versus less than one minute for the import of the \textit{RotterdamLoD12.gml} data) could perhaps be attributed to the different RAM values with the first participant having 16GB and the second 32GB\@.
However, this did not make a difference in terms of the export time for the \textit{Amsterdam.gml} dataset, reported at 5--20 minutes by both participants.
\item For the two \textit{ArcGIS Pro} tests, one user reports that the \textit{RotterdamLoD12.gml} dataset caused the software to crash when zooming and panning, meaning that results could not be compared.
All other results (\textit{Amsterdam.gml} and \textit{BuildingsLoD3.gml}) were similar
\item For the \textit{Amsterdam.gml} dataset, two out of three of the \textit{ArcGIS} testers managed to import and visualise the data, whereas the third user reported that this crashed their system.
This user also reported slower export times (1--5 minutes) for the \textit{BuildingsLoD3.gml} and \textit{RotterdamLoD12.gml} datasets (in comparison to less than a minute reported by the other two users).
Examining the hardware used, it can be noted that all three users had 16GB of RAM in their machines and dedicated graphics cards, and the only potentially significant difference between the machines is that the first two are running on an Intel i7 machine and test 3 was run on an Intel i5 machine, however the latter has a reported CPU speed (from the user) of 3.2GHz whereas the former two both use the Intel i7--8750H chip which reports a speed of between 2.2 and 4.1GHz.
Similar amounts of hard drive space were available for all three tests (270, 210 and 298GB).
Perhaps importantly it was not specified whether the hard drives were solid state --- which could be a significant factor when importing a large dataset and writing it temporarily to disc.
\item The first \textit{FZK Viewer} respondent was able to perform some timing tests on analytical tasks using the Amsterdam data, and reports that tests of Type 1 took more than 1 hour (XML Schema validation) and less than a minute (distance measurement).
However, while the first report for \textit{FZK Viewer} reported timings for some analytical tasks, the second report did not include this information.
The second \textit{FZK Viewer} report did not include time measurements for panning the \textit{BuildingsLoD3.gml} dataset.
While visualisation times for both \textit{FZK Viewer} reports were equal for the \textit{Amsterdam.gml} dataset, a marked difference could be noted when it came to zooming into and panning around the model with one user report a time of more than one hour and the other ``it is almost immediate''.
For query and inspection, the first test reports 5--20 minutes whereas the second reports ``more than one hour''.
There is no particular difference in the hardware used for testing that would account for this difference (both machines have 32GB RAM and are running Windows 10 with an Intel i7 processor).
Additionally, for the other visualisation test (rotation) very similar results are reported. 
\end{itemize}

\subsection{Writing of CityGML files}\label{sec:interoptask3}


The data exported from the tested tools \citep{noardofrancesca2020-exportsTask3}\footnote{\url{https://doi.org/10.5281/zenodo.3966915}} and delivered by participants were analysed by means of the tools described in Section~\ref{sec:mettask13}. 

The information about georeferencing is always kept unchanged, with two exceptions: \textit{novaFACTORY} for exporting \textit{amsterdam.gml} and one of the \textit{ArcGIS} tests exporting \textit{BuildingLoD3.gml} dataset.
The \textit{amsterdam.gml }dataset exported by \textit{novaFACTORY}, reports the reference system Amersfoort / RD New, EPSG:28992, Dutch national reference system for plane coordinates, instead of the EPSG:7415, which is simply associating the same CRS to the Dutch national height system ('NAP height'). Consequently, the z values of the file bounding box are different, although it is possible to change the reference system to the correct one for the export.
In one of the tests with \textit{ArcGIS} for the \textit{BuildingsLoD3.gml} dataset, the coordinates of the bounding box are slightly moved from minimum (-0.7,-0.61, 0) to (0,0,0) and maximum (70.39, 67.54, 16.7) to (69.93, 67.21, 16.7), with related consequences in the total dimension of the bounding box and area of the base surface.

In the Tables~\ref{fig:amsterdammodels}, \ref{fig:buildingsmodels} and~\ref{fig:rotterdammodels}, the differences between the statistics and computed features (related to both semantics and geometry) in the original datasets and the ones which were exported by the tested tools are represented.
In the tables on the left, the absolute number of features that are lost (in orange gradients in the tables) or generated (in red gradients in the tables) with respect to the original files are counted.
Only when the value is 0 (green) there were no changes in the files and the data exchange was successful.
These tables also include the entities and geometric elements which were not present in the original files, but appeared instead in the exported ones.

In the right part of the tables, the same values are expressed as a percentage of the whole original amount, in order to allow a more meaningful representation. 
We can understand how the interoperability is only achieved by FME and only with the \textit{Amsterdam.gml} and \textit{BuildingsLoD3.gml} datasets. It does not completely maintain the features unchanged in the \textit{RotterdamLoD12.gml}.
This would not be acceptable for use in practice.

Since a software could be tested more than once, with different results, the same software can appear more than once in the table, reporting the analysis of the delivered models.

\begin{table}[H]
	\centering
	\includegraphics[width=1\linewidth]{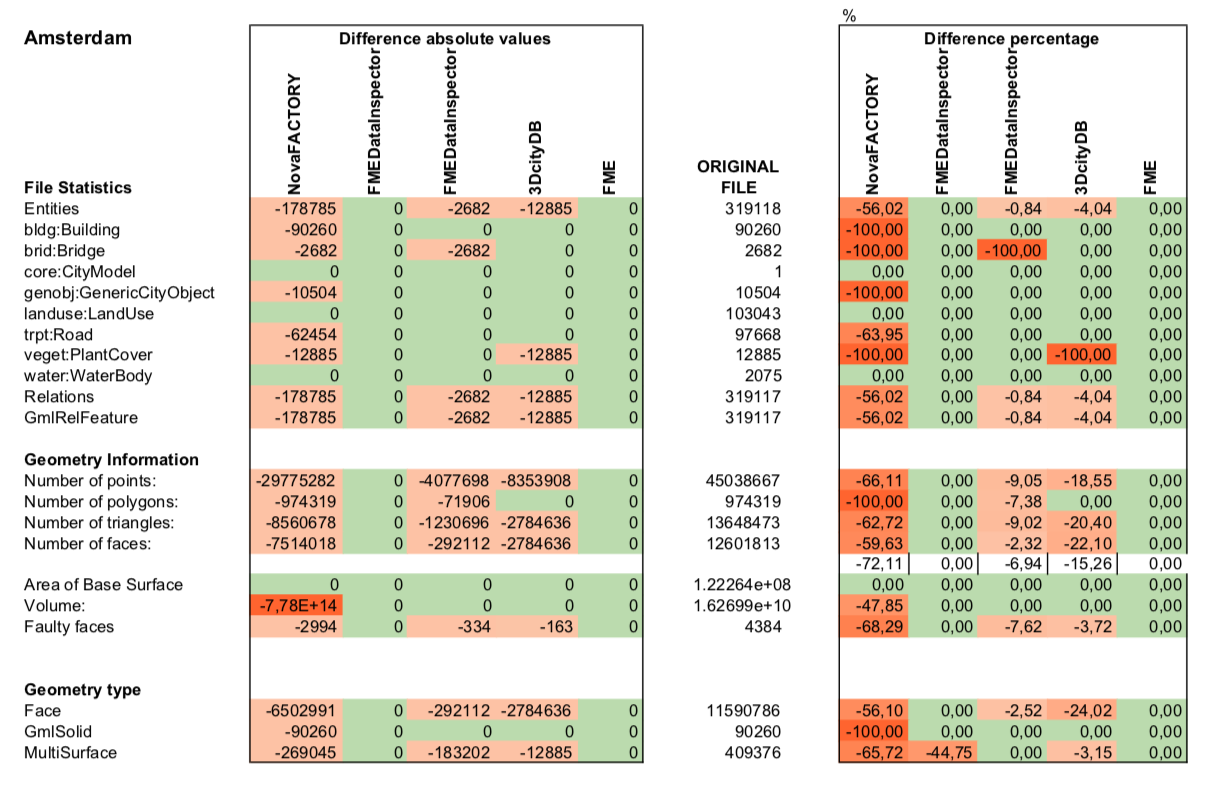}
	\caption{Comparison between the \textit{amsterdam.gml} models exported back to CityGML by participants with respect to the original one. Orange gradients are used for missing entities and green is to denote no change.}%
	\label{fig:amsterdammodels}
\end{table}

\begin{table}[H]
	\centering
	\includegraphics[width=1\linewidth]{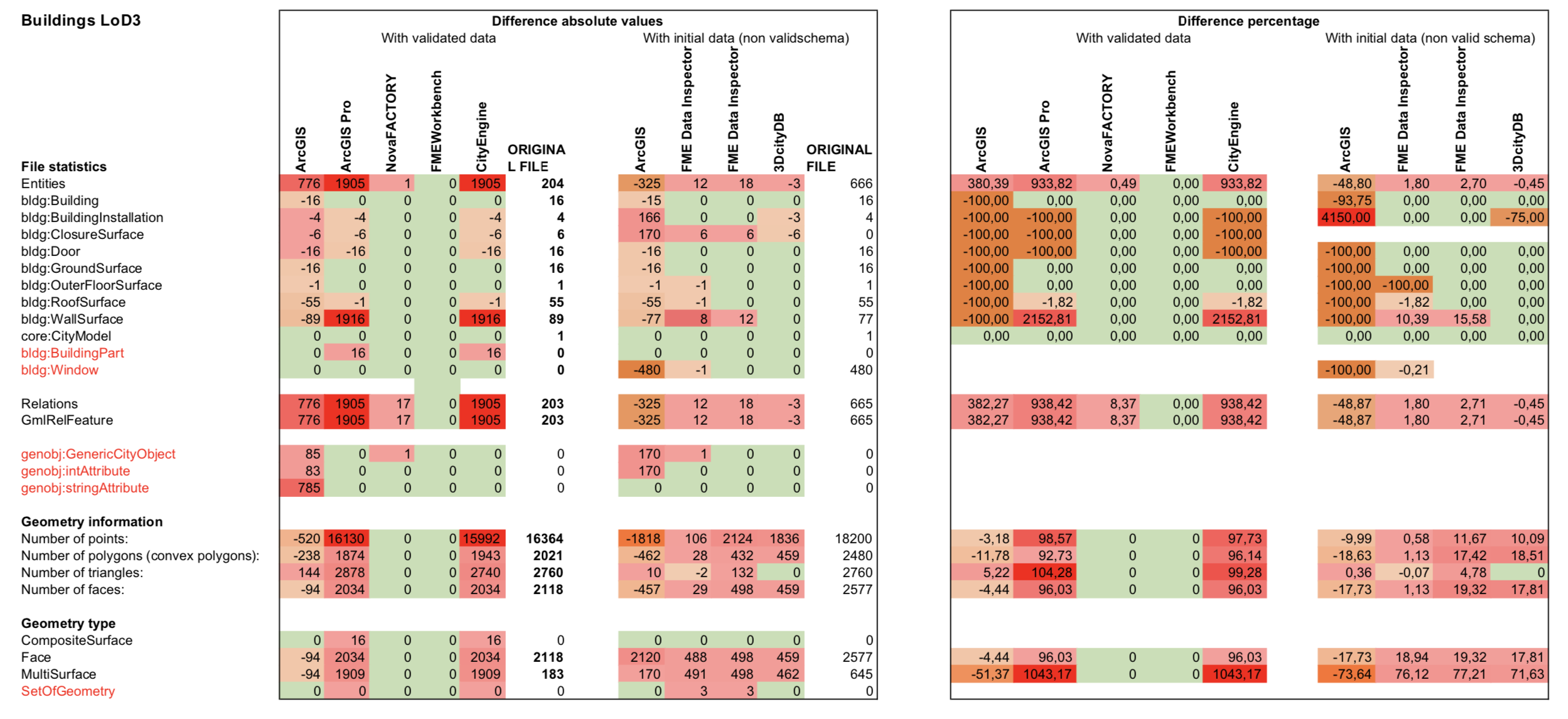}
	\caption{Comparison between the \textit{BuildingsLoD3.gml} models exported back to CityGML by participants with respect to the original one. Orange gradients are used for missing entities, red gradients for added entities and green is to denote no change.}%
	\label{fig:buildingsmodels}
\end{table}

\begin{table}[H]
	\centering
	\includegraphics[width=1\linewidth]{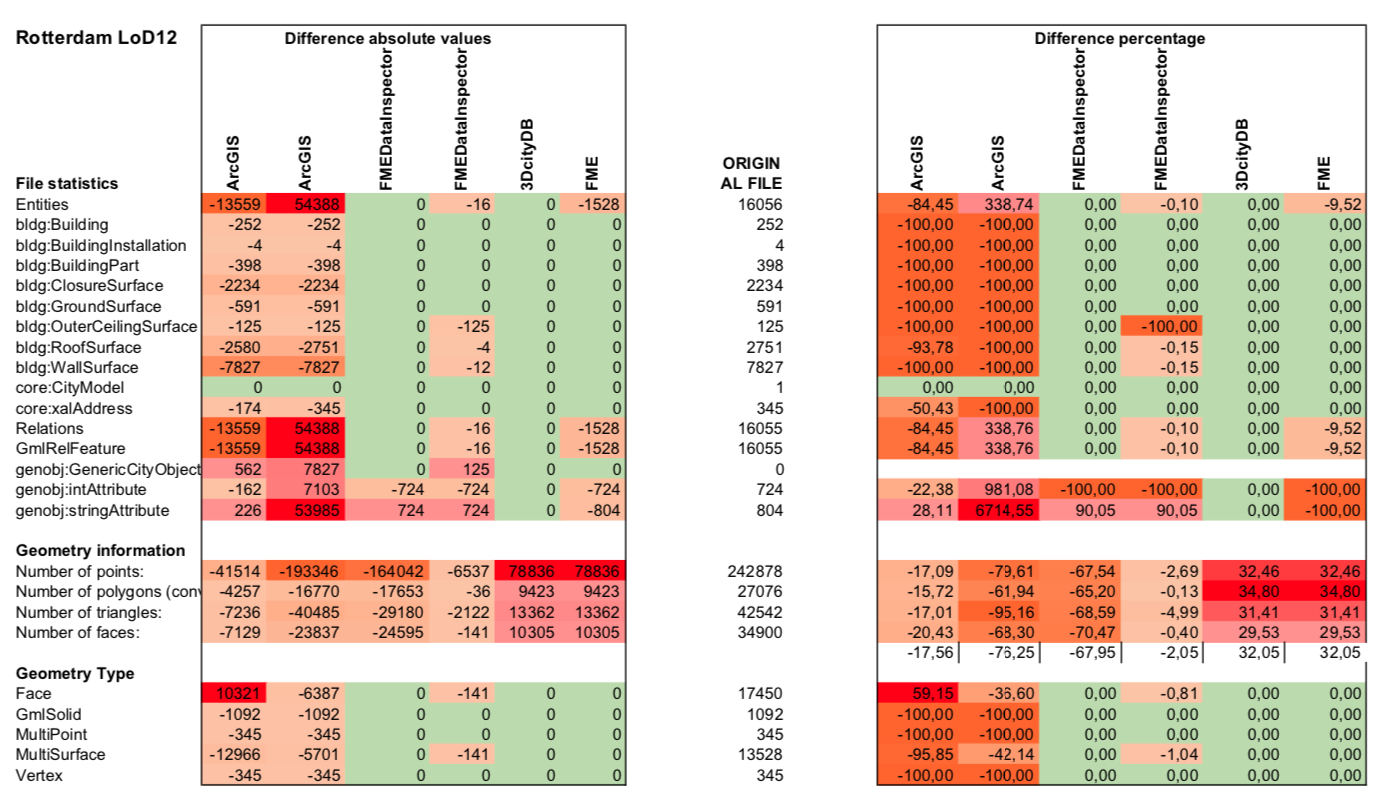}
	\caption{Comparison between the \textit{RotterdamLoD12.gml} models exported back to CityGML by participants with respect to the original one. Orange gradients are used for missing entities, red gradients for added entities and green is to denote no change.}%
	\label{fig:rotterdammodels}
\end{table}

With the \textit{Amsterdam.gml} dataset (\Reftab{fig:amsterdammodels}) a lot of entities are lost.
It can be due to the big size of the dataset, requiring very high performance by computers, which may be not able to complete the export task completely (it is probably what happened with the \textit{novaFACTORY} software).
In other cases  (\ie\ one of the \textit{FME Data Inspector} test and \textit{3DcityDB}), it is possible to notice that some specific entities are lost completely, since probably not read at all, and the related counts are affected consequently. They are the \textit{bldg:Bridges} in \textit{FME Data Inspector} and \textit{veget:PlantCover} in \textit{3DcityDB}.

In the \textit{BuildingsLoD3.gml} dataset as well some entities are completely lost.
However, in this case some other objects appear, such as \textit{bldg:WallSurface} or \textit{bldg:BuildingPart}.
It means that some internal undocumented transformation happens during import-export conversions, causing a change in entity storage or a split or duplication of objects.

A similar behaviour is also recorded for the \textit{RotterdamLoD12.gml} dataset, in which, in addition, many \textit{genobj:stringAttributes} appear unexpectedly.
Some of them derive from the change of type of some attributes previously stored as \textit{genobj:intAttributes}.
There is no apparent explanation in other cases.
Furthermore, the raise in the number of geometric elements in the \textit{3DcityDB} and \textit{FME} test remains unexplained, as well as the loss of geometries in one of the \textit{FME Data Inspector} tests, where no other changes are recorded in the number of entities.

\section{Discussion}\label{sec:discussion}

\subsection{Limited interoperability, a matter of complexity?}

Although we expected clearer patterns in the results, that could make it possible to better understand the remaining problems of interoperability in CityGML models, the only clear result is that very little interoperability is actually reached.
There are very few tools able to read the standardised datasets correctly and even fewer that are able to export them consistently.
The ability to uniquely interpret the models and to leave them consistent through the import-export phases is absolutely essential for interoperability and what it enables (data exchange, data re-use and so on).
At this stage it is not possible to trust standardised models though, even just for simple file exchange.

Furthermore, the 3D city models are expected to be usable and powerful to support analysis, problem solving and management actions.
However, the inability of tools to manage 3D information, and especially standardised 3D models in CityGML, critically hinder this potential.
This is way more apparent if looking at the discrepancy between the ability of software to correctly read and interpret the data, and their functionalities for editing, querying, analysing them.
In fact, the tools which are better at supporting those functionalities (\eg\ GIS) are not the best at interpreting the standardised datasets contents, and vice versa.
The frustration expressed informally, as reported in the introduction (Section~\ref{sec:intro}), about the gap between the users expectations and what is actually feasible (especially when they are just users and not geomatics and 3D modelling expert programmers) is therefore justified by our study.

Results were reported for 15 software packages, including both bespoke CityGML viewers but also generic GIS tools.
Support for CityGML is very mixed --- and in particular it is not directly supported (in the GML form) in perhaps the most popular open source GIS package --- \textit{QGIS}\@.
CityGML is supported in the most popular commercial package --- \textit{ArcGIS} (including Pro) but only one out of 5 users were able to display an extended 3D city model (\textit{Amsterdam.gml}) even though the display and management of an entire city model is likely to be a task that users will be interested in accomplishing.

While the georeferencing information is supported quite well by the analysed tools, problems have been reported with semantic and geometric properties.

From the gained experience, we can see how probably some difficulties in the implementation lie in the standard itself. 
The management of semantics is sometimes problematic (\eg\ loss, change), with main issues related to the management of hierarchies and other relationships.
Although being one of the most exciting possibilities in this kind of models, the lack of suitable support should compel to rethink complex relationships to make them simpler and effectively manageable.
A clear definition of how to structure entities, priorities and limits have also to be defined.
Similar considerations can be made for geometries: constrained validity rules would help a homogeneous implementation, resulting in homogeneous and consistent models.

General recommendation is therefore a collaboration of software developers with standardization institution, starting from the needs of practitioners and users.

For applications, it is often preferable to rely on a simpler model that is implemented consistently in all applications, than to dispose of a more complex one that will be implemented in fewer applications and with a lower degree of consistency.
Most applications clearly load CityGML datasets into their own internal data models: something that can be most clearly seen by the differences in the imported and exported files, which proves that the assumption that the complex data model of CityGML is directly implemented in applications is false.

The issues and challenges shown by the results of this test bench are influenced by a number of factors, among which:
\begin{itemize}
\item the complexity of CityGML and implicitly of 3D city models; a complex data model such as CityGML is required to describe and render a wide range of characteristics of the city, from both a geometrical point of view but moreover from a semantic point of view.
\item a CityGML dataset combines features and informations from both 3D graphics and geoinformation domain, which in turn causes issues to various software which are not able to handle easily this data model. Some software are tailored to more specific domains: geoinformation software (GIS, ETL, etc.) which are not always focusing on complex data structures or 3D graphics; whereas other software are more tailored for 3D graphics/visualisation, such as 3D modellers, but do not emphasize on semantics.
\item another factor which may influence the interoperability issues is the way the standard is exploited and used by users --- \eg\ how they populate the schema and which encoding they use;
\item last but not least, issues may also arise due to the way a software uses computational resources. For example, some software are more efficient in using CPU and GPU power and could handle complex CityGML data in a more efficient way, whereas other software only rely on CPU power (and may not use simultaneously more than one CPU thread) therefore causing issues when reading/writing/exporting data, which could also mean other type of data not just CityGML\@.
\end{itemize}

\subsection{The computational load}

While it is not possible to say which software package is fastest (the approach to timing used general timing categories rather than requiring the user to undertake the onerous task of time measurement), and performance will also depend on the hardware being used, we can report that none of the software packages managed to carry out the visualisation task in under 5 minutes for the \textit{Amsterdam.gml} data, although 13 packages achieved this with the \textit{RotterdamLoD12.gml} dataset and 18 for the smaller \textit{BuildingsLoD3.gml} dataset.
For CityGML export 10 software packages managed to export the data in 5 minutes or less for \textit{RotterdamLoD12.gml}, 11 for \textit{BuildingsLoD3.gml} but only 1 for \textit{Amsterdam.gml}.

A simplification of the geometry representation mechanisms could improve the related software performances, greatly affecting the time spent parsing and loading files, as well as the memory used while parsing the notoriously verbose GML geometry definitions.
Since the industry trend goes towards the direction of more and more complex and rich city models, including heavy monitoring and sensor data and so on, the need of a more agile system to store 3D city models is a urgency.

\subsection{The voluntary participation}

The fact that the inquiry is based on voluntary and completely open contribution is considered both a strength point and a limitation of this work, since it is essential to cover the investigated object in the most thorough way: as many software as possible, with as many highly experts involved, inclusion of also less expert users to test also user-friendliness. However, the limit is the incompleteness of the resulting tools review.
This has been limited by the integration of additional packages in the testing that had not been considered at the beginning.
Moreover, the tests reporting suspicious results according to the promoting team experiences, for both too good or too bad performances, were double checked with new tests or asking for clarifications.
A further issue could be the inexperience of some testers, reporting about tools behaviour in an inaccurate way.
To lower this eventuality, it was checked that all the delivered answers were described with sufficient care, whatever the level of expertise could be. Once verified this, eventual conflicting answers with respect to the tools actual potential, could indicate a deficiency in the suitability of the tool to be used by any inexpert user, which would be anyhow necessary for the models to be used in practice. The involvement of a large part of the community is also important to perceive the relevance of the topic, although dealing with a somehow hidden issue among the high-level standardization and academic communities.

\section{Conclusion}\label{sec:conclusion}

This study was designed to point out and provide evidence about the support and issues of available software for standardised information in CityGML version 2.0.
Interoperability is essential for a number of use cases, and even for merely exchange and re-use data.
Standards are supposed to be enabling such interoperability and standardization is the essential premise to the development of any integration, including the GeoBIM one.
In fact, the potential integration with other kinds of data, including Building Information Models (BIM) and respective standards, requires structures and formats to be respected reliably within datasets.
Data which are compliant to a standard are not supposed to change or vary when produced or used by different tools, otherwise, it would be almost impossible to plan, design and implement effective solutions for mapping, conversions and object transformation to other formats. 

CityGML is a popular open reference standard for 3D city models management and storage.
However, a number of issues are informally reported on using CityGML as data format, preventing the effective use of such datasets. But no systematic proof was available to be the base of future improvements in implementations, in data modelling and in the standard itself.

Possible bias in the results could be given by the eventual little expertise of participants making the tests, or by the initial inaccuracies in the provided datasets, coming from practice.
However, such datasets were validated and improved for the purpose of the benchmark, in order to limit their influence on the quality and reliability of results.
Therefore, if any of such chances happened, although the great efforts in controlling them, it would reflect additional drawbacks of the standard itself, for the little clarity about its use for the modelling of actual datasets and the difficulty in implementation, which could produce little intuitive tools.

This study shows the drawbacks of the CityGML (v. 2.0) standard and implementation, and the related difficulties, also due to the challenges to which it is intended to respond (\eg\ representation of the information regarding a huge and complex field, flexibility to multiple needs).
The outcomes are of great importance to acknowledge them officially, and to be the base for future research in the field and development of concrete solutions, such as the addition of constraints and specific guidelines, more simple ways to store geometry, better selection of useful semantics, and so on.

Considering the results of this study as an evidence, future work should be aimed at resolving the outlined issues.
For example, the test about specific kinds of geometries, how to constrain them and how to guarantee that software can import, read, use and re-export them without any change.
The same is true for semantics: which are the essential categories and relationships for the models to be useful?
How could those be simplified without losing effectiveness?
Moreover, when considering the performances related to the computational requirements, we can easily understand that the reduction of data size is urgent.

Some work is already being done to solve such issues, trying to work for less complex models offering more straightforward choices, easier to implement, for example with the introduction of CityJSON\@.
On the other hand, CityGML version 3.0 is about to be released~\citep{kutzner2020citygml}
This new version introduces many features such as a new LoD concept (based on \citet{lowner2016proposal})
extended version management system (introducing bitemporal time stamps, see
\citet{chaturvedi2017managing}),
specific classes to facilitate easier conversion to IFC, and the possibility to introduce logical spaces (as a complement to physical spaces). The latter can be used for \eg\ 3D cadastre units which enables a connection between CityGML with \eg\ the Land Administration Domain Model~\citep{ISO2012},
without introducing an ADE~\citep{sun2019utilizing}.
CityGML version 3.0 has interesting features for establishing new national standards~\citep{eriksson2020requirements},
however it is uncertain if this new standard will be able to address the technical interoperability issues evaluated in this benchmark. And there is a risk that the new features (even though interesting from several application perspectives) may even increase the complexity as highlighted in this study and therefore pose further interoperability challenges.

\section{Acknowledgements}
This work was possible thanks to the collaboration of the whole GeoBIM benchmark team (with their work as in-kind contribution to the project), all the data providers, the participants making the tests, listed in the GeoBIM benchmark website\footnote{\url{https://3d.bk.tudelft.nl/projects/geobim-benchmark/participants.html}} and the participants to the GeoBIM benchmark workshop\footnote{\url{https://3d.bk.tudelft.nl/projects/geobim-benchmark/events.html}}.

The benchmark was funded by ISPRS and EuroSDR\@. This project has also received funding from the European Union's Horizon 2020 Research \& Innovation Programme: European Research Council (ERC) under the  (grant agreement no. 677312, Urban modelling in higher dimensions) and Marie Skłodowska-Curie (grant agreement No. 707404, Multisource Spatial data Integration for smart City Applications).


\bibliographystyle{plainnat}
\bibliography{Task3finalbenchmark-rev}

\end{document}